\documentclass[11pt,a4paper]{article}
\usepackage[fleqn]{nccmath}
\usepackage{color}
\usepackage{caption}
\usepackage{graphicx}
\usepackage{subfig}
\usepackage{multirow}
\usepackage{authblk}
\usepackage[square,sort,comma,numbers]{natbib}

\begin{document}

\title{PromID: human promoter prediction by deep learning}

\author[1]{Ramzan Umarov}
\author[1]{Hiroyuki Kuwahara}
\author[1]{Yu Li}
\author[1, *]{Xin Gao}
\author[2, *]{Victor Solovyev}
\affil[1]{King Abdullah University of Science and Technology, Thuwal 23955-6900, Saudi Arabia}
\affil[2]{Softberry Inc., Mount Kisco,NY 10549, USA}
\affil[*]{To whom correspondence should be addressed.
Victor Solovyev Email:  victor@softberry.com; Xin Gao Email:xin.gao@kaust.edu.sa}
\date{}

\maketitle

\begin{abstract}
Computational identification of promoters is notoriously difficult as human genes often have unique promoter sequences  that provide regulation of transcription and interaction  with transcription initiation complex. While there are many attempts to develop computational promoter identification  methods, we  have no reliable  tool to analyze long genomic sequences. In this work we further develop our deep learning approach that was relatively successful to discriminate short promoter and non-promoter sequences. Instead of focusing on the classification accuracy, in this work we predict the exact positions of the TSS inside the genomic sequences testing every possible location. We studied human promoters to find effective regions for discrimination and built corresponding deep learning models. These models use adaptively constructed negative set which iteratively improves the models discriminative ability. The developed promoter identification models significantly outperform the previously developed  promoter prediction programs by considerably reducing the number of false positive predictions. The best model we have built has recall 0.76, precision 0.77 and MCC 0.76, while the next best tool FPROM achieved precision 0.48 and MCC 0.60 for the recall of 0.75. Our method is available at http://www.cbrc.kaust.edu.sa/PromID/.
\end{abstract}

\section{Introduction}
The high fidelity of the RNA polymerase II (pol II) transcription system is necessary for precise spatiotemporal regulation of endogenous protein expression and essential to proper development and homeostasis in eukaryotes. 
Among the key \emph{cis}-regulatory modules for RNA pol II-mediated transcription is the core promoter, which is typically situated within a DNA segment spanning from -40 bp to +40 bp relative to the transcription start site (TSS) at position +1 \cite{Kadonaga2012,Danino2015,VoNgoc2017a}.
This stretch of DNA serves as a platform on which RNA pol II and a number of auxiliary factors assemble into the transcription machinery, which is capable of integrating a range of intrinsic and extrinsic signals, to ultimately determine the proper initiation of transcription \cite{Lodish2000,Butler2002,Morris2004,Juven-Gershon2008,Kadonaga2012,Roy2015,Zabidi2015,VoNgoc2017}.
Thus, the characterization of the structure-function relation of the core promoter is crucial to unraveling the complex molecular control mechanisms underlying not just the constitutive basal expression but also the regulated expression in the RNA pol II transcription system.

Decades of \emph{in vitro} research has identified a number of functional sequence motifs for the RNA pol II core promoter \cite{Butler2002,Smale2003,Roy2015,VoNgoc2017}.
Among such functional core promoter elements, perhaps, the most well-known is the TATA box, which was, in the past, thought to be universally present in RNA pol II core promoters \cite{VoNgoc2017}.
However, the advent in genome-wide TSS detections based on high-throughput sequencing revealed that the core promoter structure is highly diverse and complex, and there are no universal core promoter elements \cite{Lenhard2012,Kadonaga2012,Roy2015,Zabidi2015,Arnold2017,VoNgoc2017}.
Indeed, recent estimates showed that only about 17\% of eukaryotic core promoters contain the TATA box \cite{Yella2017}.
More surprisingly, genome-wide structural analysis found that many core promoters do not possess any of the known core promoter elements.  
Such structural heterogeneity permits the core promoter to expand its functional repertories so as to serve as gene- and cell-type-specific transcription regulator that responds to a range of conditions; however, because of this large diversity, the design principle of the core promoter still remains largely elusive \cite{Roy2015,Arnold2017,Garieri2017,VoNgoc2017}.

The structure of the human promoter is notoriously complex and diverse.
One explanation for this is that such complex and diverse structures must be ``designed'' to properly control expression of $\sim$25,000 protein coding genes based on interactions with only $\sim$1,850 transcription factors in the human genome \cite{Maston2006}. 
Another explanation comes from a molecular evolution study which discovered substantially accelerated rates of evolution in primate promoters compared with other mammalian promoters \cite{Taylor2006}.
This rapid primate promoter evolution was found to be comparable to the neutral substitution rate, suggesting that primate promoters have weak selective constraints, and this suggestion can also explain highly complex and diverse structures in the human promoter.
In any case, a better understanding of the structure-function relation of the human promoter has particularly important implications as some genetic variants in such noncoding regions are associated with rare Mendelian diseases 
\cite{Edwards2013,Rojano2018}.
Furthermore, some cancer cells are associated with somatic mutations in promoter regions \cite{Vinagre2013,Fredriksson2017}.
In order to gain insights into what types of genetic variations can cause aberrant expression leading to human diseases, it is crucial to accurately predict the locations of human promoters and to understand their structural patterns.

Here, we introduce PromID, a novel machine learning-based approach for the prediction of human RNA pol II core promoters.
Taking advantage of big promoter collection with experimentally validated TSSs \cite{dreos2016eukaryotic} generated by modern high-throughput  techniques, PromID builds a deep learning model using sequence data as its input.
To avoid bias based on prior knowledge about promoter loci (e.g, sequences with known core promoter elements and high density of CpG dinucleotides) PromID does not use predefined features; but rather it attempts to discover sequence features and learn salient patterns of the human promoter solely from the training set.
This is important especially in the prediction of human promoters since the structural features of many promoters are still unknown \cite{Maston2006,Roy2015}.
We previously developed a similar convolutional neural network-based algorithm for the prediction of core promoter locations in several model organisms \cite{Umarov2017}. While this method was able to outperform previously developed promoter prediction methods \cite{Umarov2017}, its false positive rate was not adequate enough to ensure the accurate detection of promoters on long genomic sequences.  
PromID was developed to chiefly alleviate this limitation and to focus more on the promoter prediction on longer sequences.
Specifically, to reduce the false positive rate, we designed PromID to adaptively and iteratively train the predictor by changing the distribution of samples in the training set based on false-positive errors it made in the previous iteration. 
By increasing the weight of difficult non-promoter sequences in the training set, we can force the predictor to learn promoter patterns to rule out such sequences.   
To evaluate the performance  of PromID, we compared our neural network model with publicly available tools for the human promoter prediction task.
We found that the PromID outperformed the other predictors and achieved much smaller error-per-1000-bp rate than the others.
Our results demonstrate the usefulness of PromID for the human promoter prediction on long genomic sequences and suggest its potential value as a tool to gain insights into the design principle for the human core promoters.

\section{MATERIALS AND METHODS}

\begin{figure*}[t]
\begin{center}
\includegraphics[width=125mm]{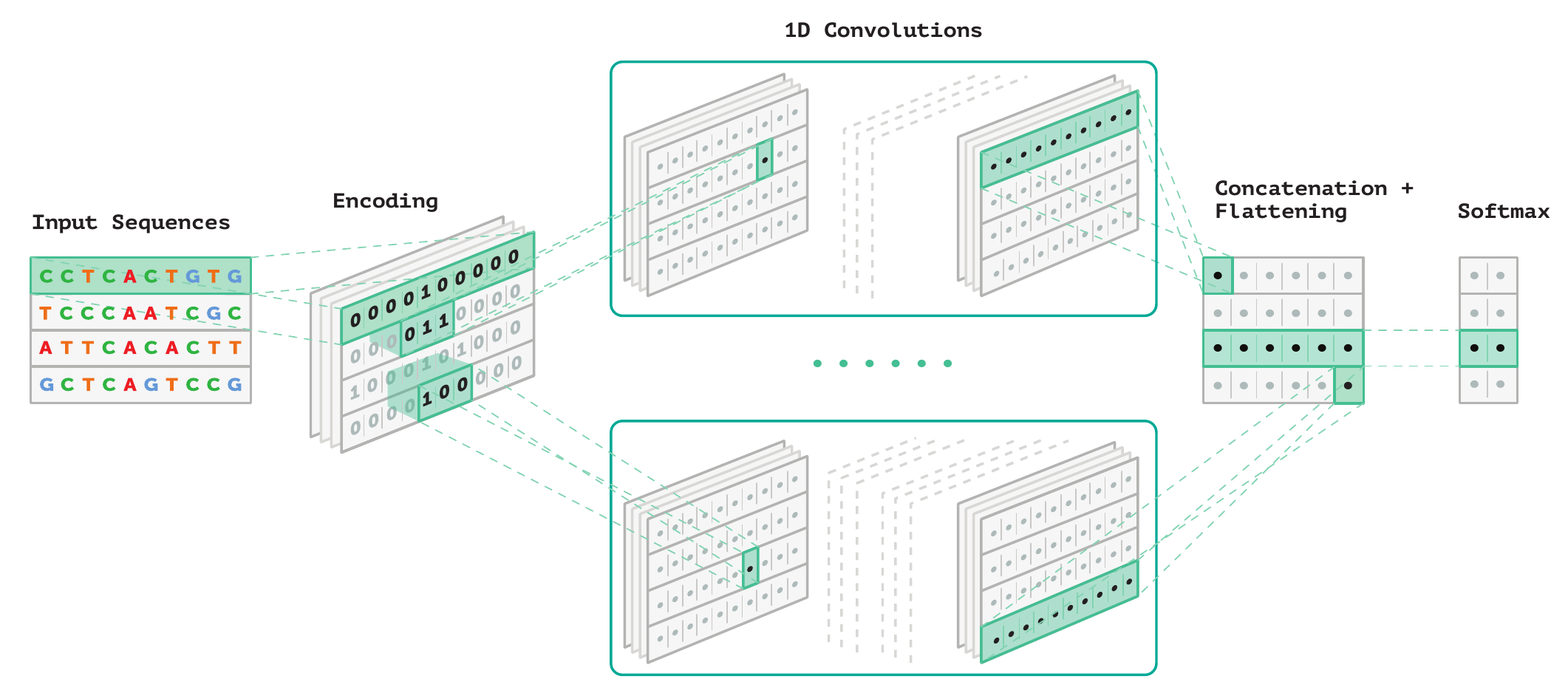}
\end{center}
\caption{Deep learning model architecture that was used in building promoter models of PromID (see text for its description).}
\label{fig:Architecture}
\end{figure*}

\subsection{Datasets}
Our models are trained using human promoter sequences extracted from the EPDnew database \cite{dreos2016eukaryotic}. 
The EPD database is an annotated non-redundant collection of eukaryotic POL II promoters,
for which the transcription start site has been determined experimentally.
The authors of the EPDnew database have demonstrated its higher quality over the ENSEMBL-derived \cite{aken2016ensembl} human
promoter set \cite{dreos2012epd}.

In this study we downloaded  16455 genomic sequences (from -5000 bp to +5000 bp,  where +1 is a TSS position) containing human promoters from the EPD database. We used 90\% of the sequences for training and 10\%  for testing. Positive and negative sets were extracted from the training set. A promoter region of a given size around the known TSS is considered to be a positive sequence. A negative sequence is the one outside the promoter region which does not contain a known TSS. Initially, the negative set had the same size as the positive one and consisted of randomly picked negative sequences. 

\subsection{Deep neural network model}
To classify promoters from non-promoters deep neural networks were used. 
The data is read in the fasta format and then transformed using one hot encoding. 
This encoding uses a vector of size 4 to represent each nucleotide.
\textit{A} is encoded as (1 0 0 0), \textit{T} is encoded as (0 1 0 0), \textit{G} is encoded as (0 0 1 0),
and \textit{C} is encoded as (0 0 0 1). For each input nucleotide sequence, the one-hot
encoding would produce an L by 4 matrix, where L is the length of analyzed sequence window.

Our architecture consists of several Convolutional Neural Networks (CNN) layers
which are in parallel (Figure \ref{fig:Architecture}). This means that they all have access to the original input.
The filter lengths are different for each layer and they are able to represent different 
promoter elements. We also have a special layer with filter length just one which is 
able to capture GC content of the sequence  
which is known to be higher in a promoter region \cite{fenouil2012cpg}.
This layer is the only one followed by average pooling while all others use maximum pooling since we only care about the count of \textit{G} and \textit{C} nucleotides, not their positions inside a promoter. The CNN layers are concatenated, flattened and fed into a softmax layer. We do not have a fully connected dense layer as we noticed it actually reduces the predictive power of our model. 

The softmax layer has two neurons which represent an input sequence being promoter ($p_{p}$) or non-promoter ($p_{np}$). The final score produced by our model is calculated as follows:
\begin{ceqn}
\begin{equation*}
Score=\frac{(p_{p} - p_{np} + 1)}{2}.
\label{eq:SC}
\end{equation*}
\end{ceqn}
This score has values in the range from 0 to 1 and is our approximation of probability that an input is a promoter. 

We use three methods to improve the generalization capability of our model.  The first method is weight decay,
which effectively limits the number of free parameters in the model 
to avoid overfitting. The deep neural network model is likely to reproduce the noise
by using extreme weights. Introducing weight decay makes it possible to regularize the cost function by penalizing large weights. The second method is dropout \cite{srivastava2014dropout}. The main idea of this technique is to randomly set some nodes of neural network to zero during  training to prevent co-dependency amongst each other. The third method is batch normalization \cite{ioffe2015batch}. This approach
extends the idea of normalization by making it a part of the model architecture. It reduces overfitting because it has a regularization effect similar to dropout. Also, we can use higher learning rates because batch normalization makes sure that there is no activation that is too high or too low.

Our model uses the Adam optimization algorithm to train the weights \cite{kingma2014adam}, 
which is an improved version of stochastic gradient descent.  The method computes individual adaptive learning rates for different parameters from estimates of first and second moments of the gradients.   Adam realizes the benefits of both AdaGrad and RMSProp. We use TensorFlow \cite{abadi2016tensorflow} as the framework to construct the deep neural network. The training was performed on a workstation with two 980 GTX GPUs and took on average 3 hours.

\subsection{Classification procedure}

\begin{figure*}[t]
\centering
\subfloat[Promoters with the TATA box. Corresponding genes from top left clockwise - COCH\_1, CCL5\_1, KL\_1, SEMG1\_1.]{%
  \includegraphics[clip,width=125mm]{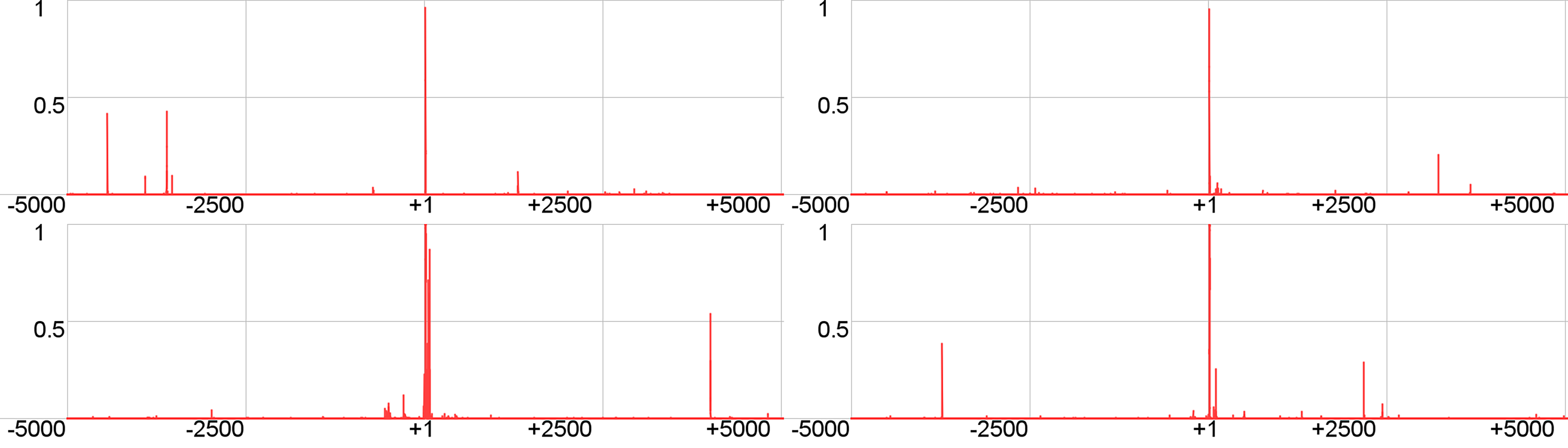}%
}
\\
\subfloat[Promoters without the TATA box. Corresponding genes from top left clockwise - FAM135A\_1, ASCC3\_1, CR1L\_1, AP3D1\_1.]{%
  \includegraphics[clip,width=125mm]{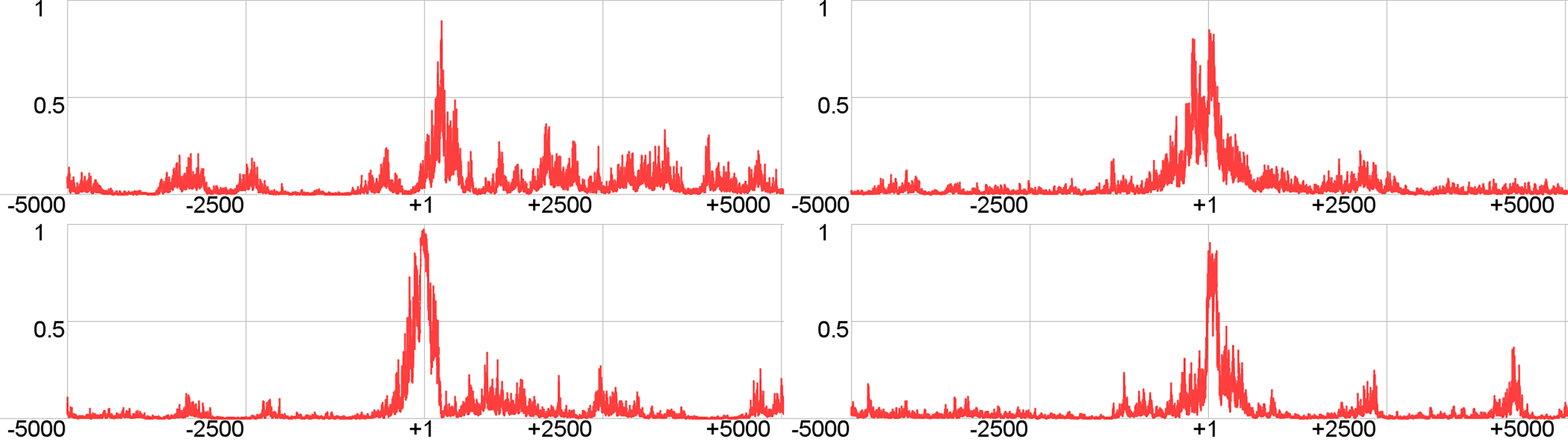}%
}
\caption{Scoring landscapes constructed by our model. True TSS is at position +1.}
\label{fig:Landscape}
\end{figure*}

We train the model using positive and negative set which consists of relatively 
short sequences with a fixed length.  As our model accepts sequences of certain length as input we apply a sliding
window approach to analyze long genomic sequences. This window is moved across the sequence
and at each position the subsequence is fed into our model. The model gives 
us a score from 0 to 1 that represents the likelihood that a subsequence
is a promoter region. If we plot these promoter scores, we will receive a scoring
landscape of our model, see Figure \ref{fig:Landscape}.

If the value of the score of a sliding window is above the threshold it is predicted as a start 
of a promoter region. In practice, we construct two deep learning models
 - one is for identification of promoter sequences with the TATA box and
one for promoters without it. We use the TATA box weight matrix \cite{bucher1990weight} to
distinguish sequences with the TATA box and without one.
Promoters can be predicted much more accurately if they have the TATA box,
that is why we firstly apply the model trained specifically for the promoters with the TATA box (TATA+ model).
Next, we apply the model trained with the promoter sequences without the TATA box (TATA- model).
We account the second model  predictions that are not too close  to the first
model predictions, the prediction are required to be at least 1000 bp 
apart. We combine their output to make the final decision
about promoter region position.  TSS is then considered to be at a certain position 
inside the promoter region. For example, if our sliding window has length 600 bp and the positive set 
was extracted from -200 bp to +400 bp, then the TSS will be located at position 201
inside the predicted promoter region.  

\subsection{Negative set construction}

When constructing the prediction model to classify promoters we need 
to choose what sequences to use for non-promoters. This problem 
is very important because it affects what features our model
will use to separate the two classes. For example, suppose we choose random 
DNA sequences which we are sure are not promoters for the negative set.
In this case, a very small number of them will have TATA motif at 
the specific position. Then the neural network model will just use this one 
feature to achieve almost perfect separation between the two classes. 
When applying such a model to real world data, the sensitivity will
be high however there will be a lot of false positives. Any sequence
with a TATA motif at the specific position will most likely be classified as a
promoter.  Simply increasing the negative set size is not an effective solution as well,
because firstly our data becomes unbalanced and secondly, there
will be a big chance that neural networks will be stuck at some
local minimum as in the case considered above. There are not many sequences in
the negative set that will have a good scoring TATA motif at the specific position, 
which makes our network likely to derive its recognition model heavily based 
on this single discriminating feature.         

\begin{figure}
\begin{center}
\includegraphics[width=70mm]{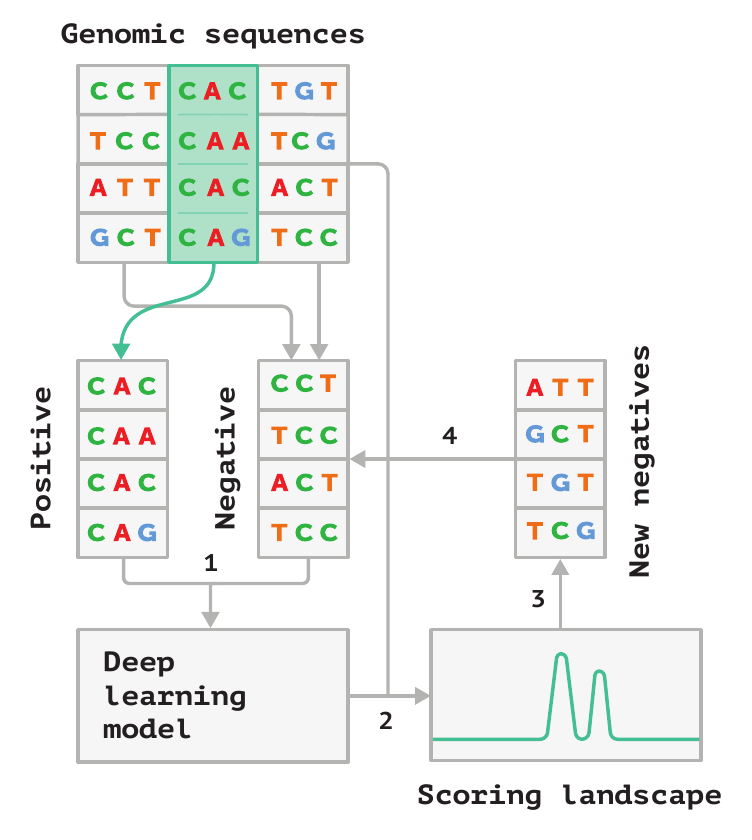}
\end{center}
\caption{Proposed training pipeline. See text for the description of each step.}
\label{fig:pipeline}
\end{figure}

To resolve these issues we propose an iterative approach described below. Firstly, we choose 
a negative set randomly. Then we repeat the following steps:
\begin{enumerate}
\item We train a model with the current negative set.
\item The model is applied to the dataset
with long sequences and false positives are recorded. 
\item  A subset of  false positives with the highest scores given to
them by our model (the ones that are most similar to the true promoters) from each long sequence are chosen
for the new negative set.
\item A new negative set is then constructed by merging the previous one with 
the new false positives. 
\end{enumerate}
This procedure is repeated until there are only  a  few false
positives found processing the training set in step 2. These steps are illustrated in Figure \ref{fig:pipeline}.
Such a procedure constructs a difficult negative set which helps to force our neural
network to learn deeper and less obvious features to recognize a promoter sequence.

\subsection{Selecting length and location of input}
\begin{figure}
\begin{center}
\includegraphics[width=90mm]{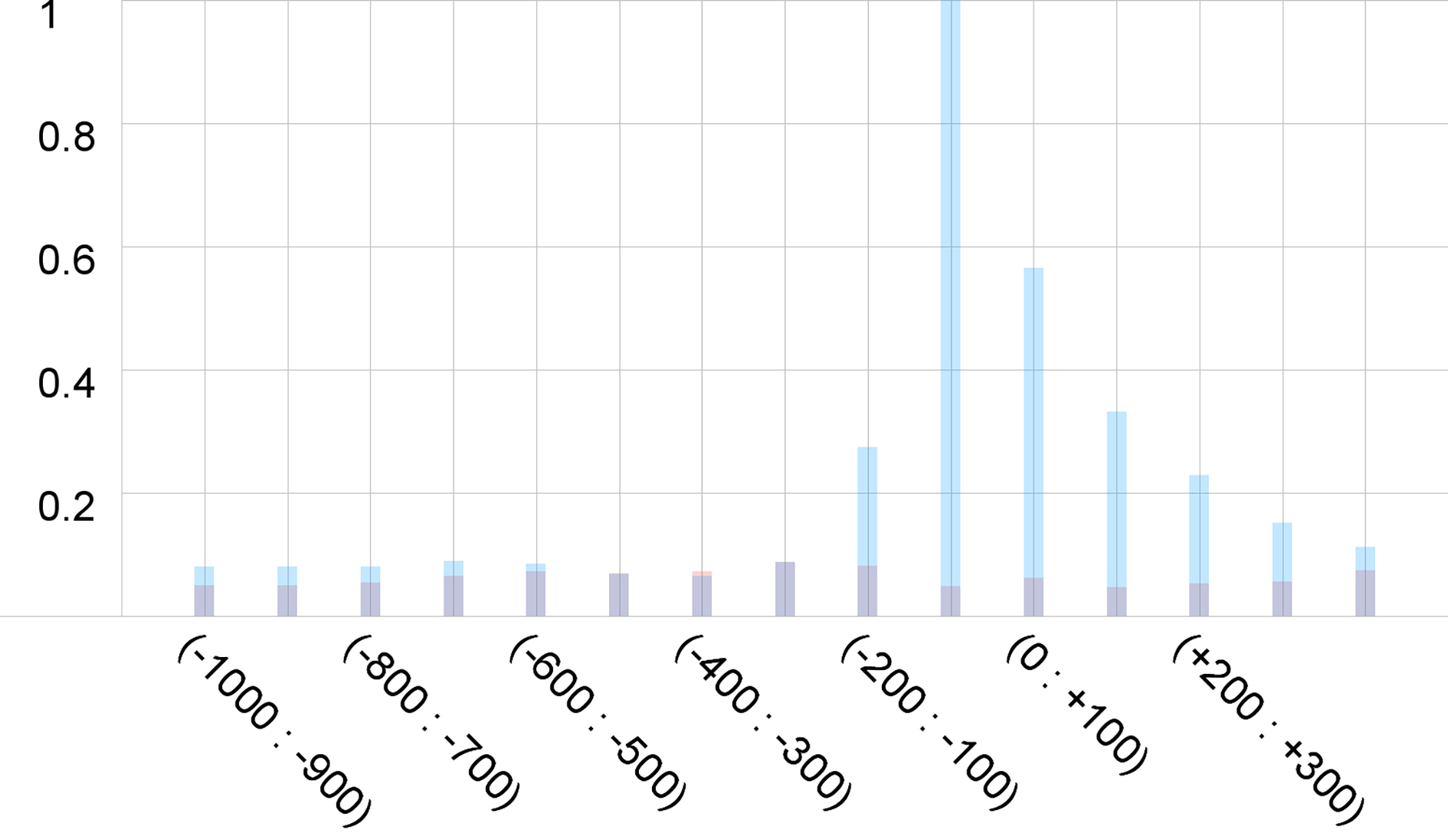}
\end{center}
\caption{Influence of different regions inside the promoter on the final score produced by the deep learning model. Blue color represents decrease of the score after random substitution and red color shows its increase. }
\label{fig:An_fig}
\end{figure}
We need to chose what part of a promoter region to feed into our model for training. 
In our previous work on promoter identification we used region from -200 bp to +50 bp to extract promoter features.  
Since multiple transcription start points \cite{wang2017punctilious, dreos2016eukaryotic} often significantly enlarge
potential gene promoter regions, in this work we decided to create  a promoter model 
using a much wider region from -1000 bp to +500 bp and then apply our random substitution procedure
to study the location of sequence elements affecting the promoter prediction performance and potentially narrow the region down.  
The random substitution procedure works as follows. We have a window of size 100 which we move along each sequence with step size 100. At each position
we replace the nucleotides with random 100 nucleotides and calculate new promoter score for the modified sequence.
The difference between the original score and the new one is recorded and reported for each position (Figure \ref{fig:An_fig}).

We noticed that the region from -200 bp to +400 bp has the most significant effect on the score predicted by our model and this is why it was used to train our 
final model.  

\subsection{Performance measures}
In order to evaluate our method and to objectively compare predictions by our models and the other promoter identification  methods, we measured performance using Recall, Precision, and Correlation Coefficient (CC):
\begin{eqnarray*}
\label{eq:CC}
&Recall=\frac{TP}{TP + FN}, \\
&Precision=\frac{TP}{TP + FP}, \\
&CC=\frac{TP\times TN - FP\times FN}{\sqrt{(TP + FP)(TP+FN)(TN+FP)(TN+FN)}}.
\end{eqnarray*}
If we predict a promoter with the TSS  which is closer to the known TSS than the allowed margin for error (500 bp) then this prediction is counted as a TP. If there is no prediction in the area from -500 bp to +500 bp of the known TSS then we count this case as a FN. Any prediction outside the region from -500 bp to +500 bp of some TSS is counted as a FP. The same rule is applied for performance evaluation of all the tested promoter prediction programs.
Also, we used  two accuracy measures that are useful to evaluate the performance of promoter prediction tools when analyzing long genomic sequences:  the average prediction error per correctly  predicted TSS (FP/TP) and the average prediction error per 1000 bp.

\section{RESULTS}

\subsection{Comparison on predictive performance}

\begin{table*}[t]
\tiny
\centering
\caption{Predictive power of the different models trained in this study. The comparison includes different input regions,
one model versus two models, and usage of the TATA box matrix.}
\label{tab:Perfomance1}
\begin{tabular}{|c|l|l|l|l|l|l|l|l|l|}
\hline
\multicolumn{1}{|l|}{}            & \multicolumn{3}{c|}{Recall} & \multicolumn{3}{c|}{Precision} & \multicolumn{3}{c|}{MCC} \\ \hline
                                  & TATA+   & TATA-   & BOTH    & TATA+    & TATA-    & BOTH     & TATA+  & TATA-  & BOTH   \\ \hline
\shortstack{1 model\\{[}-1000 +500{]}}          & 0.556   & 0.791   & 0.761   & 0.635    & 0.730     & 0.720     & 0.594  & 0.760   & 0.740   \\ \hline
\shortstack{2 model\\{[}-1000 +500{]}}          & \textbf{0.725}   & \textbf{0.813}   & \textbf{0.802}   & 0.602    & 0.641    & 0.636    & 0.661  & 0.722  & 0.714  \\ \hline
\shortstack{2 model matrix\\{[}-1000 +500{]}}   & \textbf{0.725}   & 0.806   & 0.796   & 0.612    & 0.635    & 0.633    & 0.666  & 0.716  & 0.709 \\ \hline
\shortstack{1 model\\{[}-200 +400{]}}           & 0.531   & 0.766& 0.736& 0.714& \textbf{0.786}&\textbf{ 0.779}& 0.616& \textbf{0.776}& 0.757  \\ \hline
\shortstack{2 model\\{[}-200 +400{]}}           & 0.628   & 0.773& 0.755& \textbf{0.722}& 0.775& 0.769& \textbf{0.673}& 0.774& \textbf{0.762}        \\ \hline
\shortstack{2 model matrix\\{[}-200 +400{]}}    & 0.628& 0.770& 0.752& 0.714& 0.775& 0.768& 0.670& 0.773& 0.760        \\ \hline
\shortstack{1 model\\{[}-200 +50{]}}             & 0.411   & 0.744   & 0.702   & 0.612    & 0.714    & 0.706    & 0.501  & 0.729  & 0.704        \\ \hline
\shortstack{2 model\\{[}-200 +50{]}}             & 0.604   & 0.766   & 0.745   & 0.576    & 0.641    & 0.634    & 0.590   & 0.701  & 0.687        \\ \hline
\shortstack{2 model matrix\\{[}-200 +50{]}}      & 0.609   & 0.762   & 0.742   & 0.565    & 0.613    & 0.608    & 0.586  & 0.683  & 0.672        \\ \hline
\end{tabular}
\end{table*}

The performance of different models on the EPD data testing set is shown in Table \ref{tab:Perfomance1}. We compared the effect of using different lengths of the sequence window and using one model versus using two models - one for TATA box promoters and one for promoters without it. We can see from this table that reducing the input sequence length from 1500 bp to 600 bp has no negative effect on the predictive performance, instead there is an improvement in MCC from 0.740 to 0.757.  However, reducing further to 250 bp reduces MCC considerably to 0.704. This motivated  us  to select the -200 bp to +400 bp region for feature extraction in our method.  
As expected, using two models improves how well we can predict promoters with the TATA box since we have a dedicated model for these sequences. We can see improvement in recall for promoters with the TATA box for all model lengths. However, since in the EPD data set the number of the TATA box promoters is rather small (about 12\%) the total performance slightly decreases because the TATA+ model introduces extra false positives in non-promoter genomic sequences.

We also noticed that simply applying the TATA+ and TATA- models sequentially to the sub-sequence is slightly better compared to using the weight matrix to distinguish the promoters. Even though the results were a little bit worse, we still kept an option in our tool to use the weight matrix if desired by the user.   

Using two models might be useful when the user of PromID  studies mostly  promoters with the TATA box or  in the future cases where we can add promoter specific features to the set of analyzed promoter characteristics. Moreover,  true positive predictions by the TATA+ model are very close to the true TSS with the difference in position being often less then 5 bp. However, if the user wants to minimize false positives he can use one model version which is also two times faster. 

We also compared our method to all the promoter identification tools we could obtain. A number of promoter prediction methods have been proposed. TSSW \cite{salamov1997gene} uses a linear discriminant function combining a TATA box score, triplet preferences around the TSS, hexamer preferences  and potential transcription factor binding
sites. It has shown good results in a review paper by Fickett \cite{fickett1997eukaryotic}. FPROM was created by extending the TSSW program feature set which resulted in significant improvement over TSSW and other promoter recognition software \cite{solovyev2003promh}. Promoter2.0 \cite{knudsen1999promoter2} extracted promoter elements from DNA sequences and used ANN to distinguish promoters from non-promoters based on these features. DragonGSF \citep{bajic2003dragon} also used ANN as a part of its design and considered GC content and the concept of CpG islands for promoter recognition. 

Our previous promoter recognition software, PromCNN achieved good classification performance in discriminating between short promoter and non-promoter sequences \cite{umarov2017recognition}. Very recently PromCNN was outperformed by \citep{qian2018improved} improving accuracy by about 7\%. However, as in \cite{umarov2017recognition}, they focused on the classification performance of short sequences, instead of promoter identification given a long genomic sequence. The latter is a much more difficult problem to tackle because of the high risk of having a large number of false positives. We could not compare our new method to theirs because they did not provide a web server or a tool that would accept long genomic sequences as an input.  

Some of the tools we came across are not available anymore. There are also tools that require extra information besides sequence data as an input. Thus here we compared our method with the following methods: PromCNN, TSSW, FPROM, Promoter 2.0. The results are shown in Table \ref{tab:Performance}. We noticed that all these  programs  have high recall and mostly very low precision. Thus we tried to modify their parameters if possible to reduce the number of false positives. This was beneficial to FPROM by increasing its MCC from 0.446 to 0.598 and to PromCNN for which MCC was increased by 0.174.

\begin{figure}
\begin{center}
\includegraphics[width=60mm]{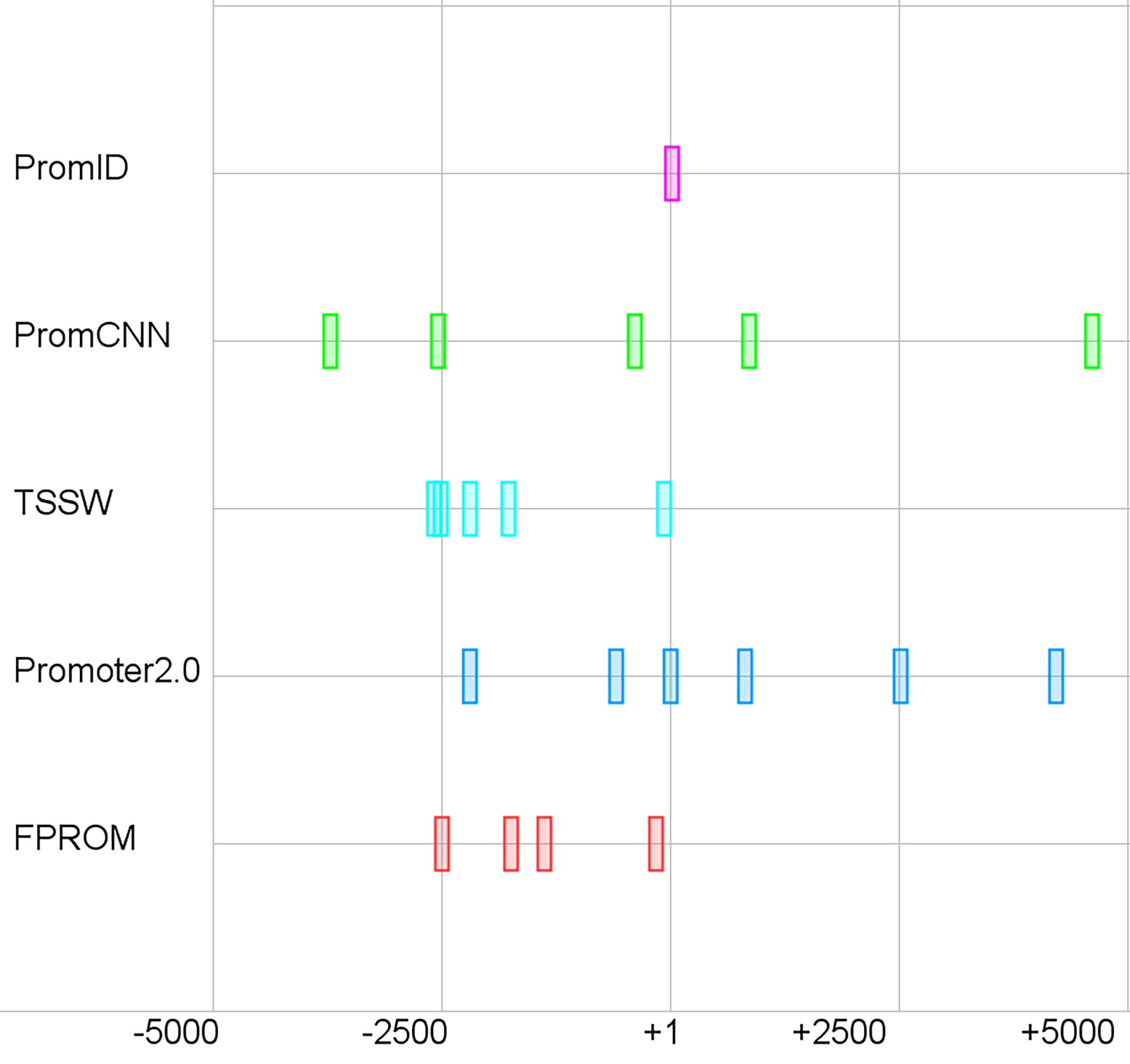}
\end{center}
\caption{Predictions of different promoter recognition tools on the promoter of UBE3D\_1 gene. True TSS is at position +1. }
\label{fig:profile}
\end{figure}

Regardless of the parameters tested, PromID significantly outperforms the competitors that were tested which showed relatively good performance in previously published papers \cite{bajic2006performance, fickett1997eukaryotic}. For example, PromID using the model trained on the [-200 +400] region has precision and MCC  higher than the best competing tool, FPROM, by 0.291 and 0.164 for the similar recall of 0.749. Figure \ref{fig:profile} shows an example of the predictions made by the different promoter prediction programs on the sequence containing the promoter of the UBE3D\_1 gene. We can see that PromID makes much less false positive predictions while still successfully finding the true TSS. 

\begin{table*}[]
\tiny
\centering
\caption{Comparison of the average performance of the different promoter prediction methods. Program runs without marginal predictions are marked with asterisk.}
\label{tab:Performance}
\begin{tabular}{|c|l|l|l|l|l|l|l|l|l|}
\hline
																																&       & PromID& PromCNN & PromCNN*& FPROM  & FPROM* & TSSW  & Promoter2  \\ \hline
\multirow{3}{*}{Recall}  																				& TATA+ & 0.628 &  0.884  & 0.700   & \textbf{0.908}  & 0.647  & 0.691 & 0.845       \\ \cline{2-9} 
																																& TATA- & 0.773 &  \textbf{0.948}  & 0.889   & 0.868  & 0.764  & 0.775 & 0.810       \\ \cline{2-9} 
																																& BOTH  &  0.755 &  \textbf{0.940}  & 0.865   & 0.873  & 0.749  & 0.764 & 0.814       \\ \hline
\multirow{3}{*}{Precision}  																		& TATA+ & \textbf{0.722} &  0.118  & 0.242   & 0.236  & 0.491  & 0.252 & 0.107       \\ \cline{2-9} 
																																& TATA- & \textbf{0.775} &  0.127  & 0.320   & 0.227  & 0.476  & 0.259 & 0.104       \\ \cline{2-9} 
																																& BOTH  & \textbf{0.769} &  0.126  & 0.310   & 0.228  & 0.478  & 0.258 & 0.105       \\ \hline
\multirow{3}{*}{MCC}  																					& TATA+ & \textbf{0.673} &  0.323  & 0.411   & 0.463  & 0.564  & 0.417 & 0.301       \\ \cline{2-9} 
																																& TATA- & \textbf{0.774} &  0.347  & 0.534   & 0.444  & 0.603  & 0.448 & 0.291      \\ \cline{2-9} 
																																& BOTH  & \textbf{0.762} &  0.344  & 0.518   & 0.446  & 0.598  & 0.444 & 0.292      \\ \hline
\multirow{3}{*}{\shortstack{Error per \\ correct} }             & TATA+ & \textbf{0.385} &  7.464  & 3.138   & 3.234  & 1.037  & 2.965 & 8.349      \\ \cline{2-9} 
																																& TATA- & \textbf{0.290} &  6.885  & 2.121   & 3.403  & 1.099  & 2.857 & 8.581       \\ \cline{2-9} 
																															  & BOTH  & \textbf{0.300} &  6.953  & 2.225   & 3.381  & 1.092  & 2.869 & 8.551       \\ \hline
\multirow{3}{*}{\shortstack{Error per \\ 1000 bp} } 						& TATA+ & \textbf{0.024} &  0.660  & 0.220   & 0.294  & 0.067  & 0.205 & 0.706       \\ \cline{2-9} 
																																& TATA- &\textbf{0.022} &  0.653  & 0.189   & 0.295  & 0.084  & 0.221 & 0.695       \\ \cline{2-9} 
																																& BOTH  & \textbf{0.023} &  0.654  & 0.192   & 0.295  & 0.082  & 0.219 & 0.696      \\ \hline
\end{tabular}
\end{table*}

\subsection{Analyzing the learned model}
\begin{figure}
\begin{center}
\includegraphics[width=70mm]{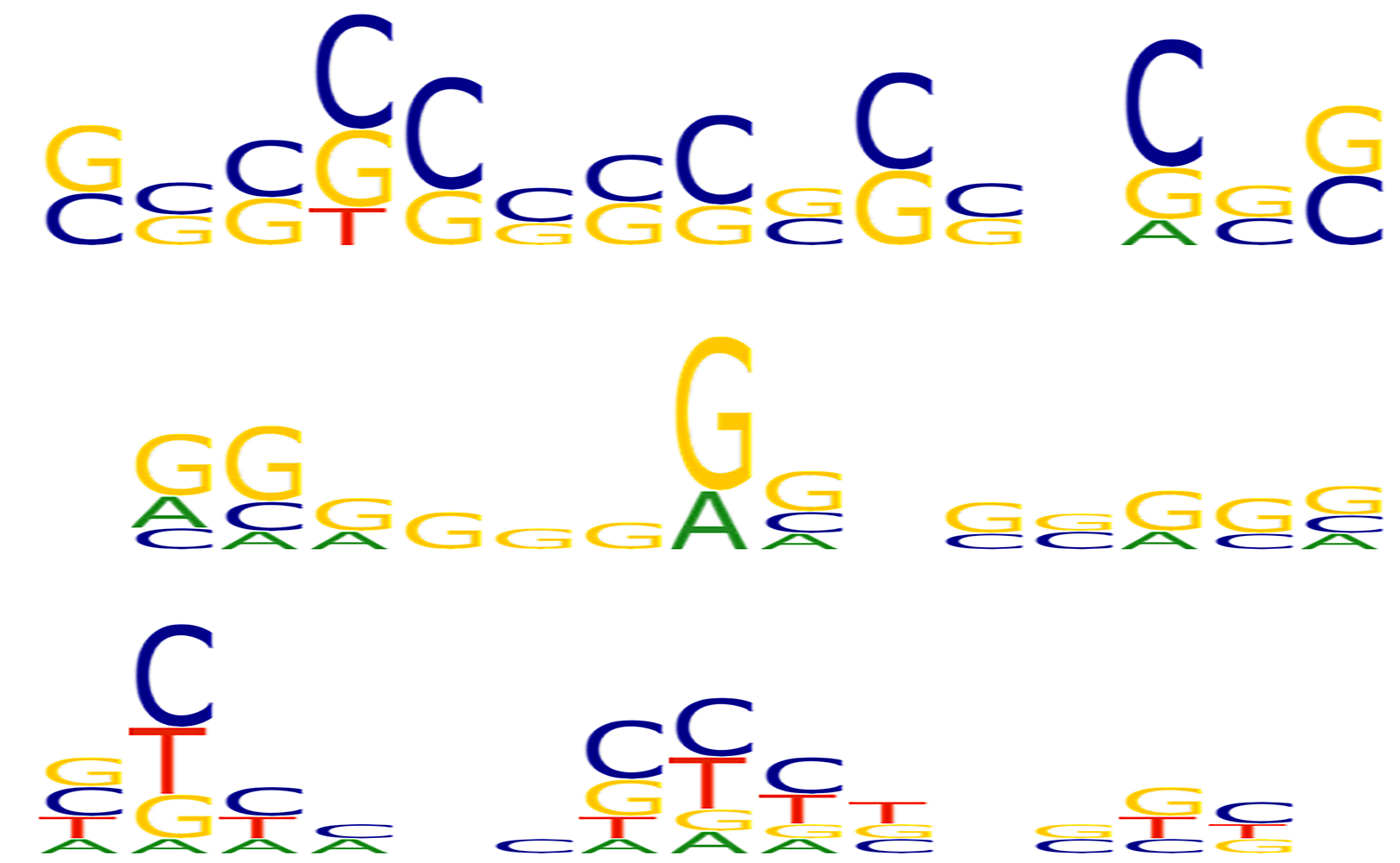}
\end{center}
\caption{Sequence logo of the most important 15-mers identified by our model. }
\label{fig:logo15}
\end{figure}

It is well-known that the models trained by neural networks are hard to interpret. We tried to overcome this limitation by visualizing the trained convolutional filters. The maximum filter length we use is 15 which is the length of the TATA box weight matrix. Thus we decided to find the most important 15-mers identified by our model. We found the top 1000 most influential 15-mers and grouped them into three groups using k-means clustering algorithm. The sequence logo for these three groups is shown in Figure \ref{fig:logo15}. The top three most important motifs were \textit{CCCAGGACCATGTCT}, \textit{GCTAGGTTGTTATGT}, \textit{GTTCCCGGCCGGTGC}, which all contain GC rich subsequences that are well known characteristics of eukaryotic promoters \cite{fenouil2012cpg}. 

\begin{figure*}
\begin{center}
\includegraphics[width=125mm]{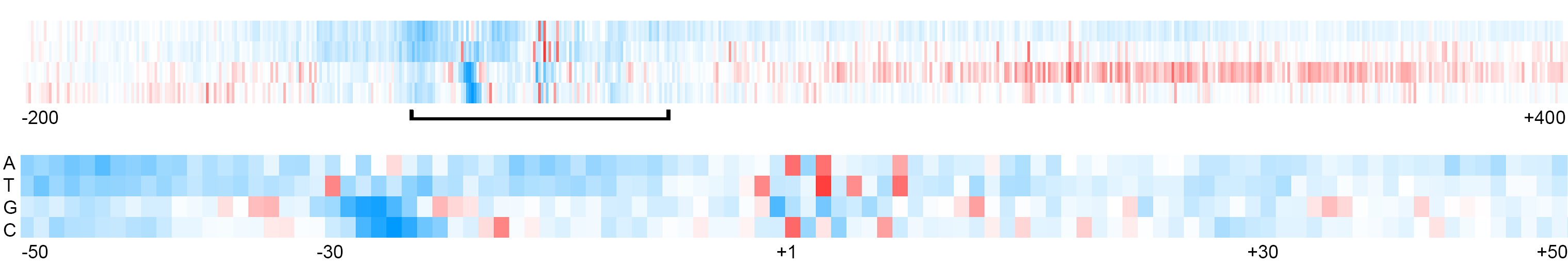}
\end{center}
\caption{Saliency map for the region from -200 bp to +400 bp. Red score represents increase of the score produced by our model and blue shows its decrease.}
\label{fig:sal}
\end{figure*}

\begin{figure}
\begin{center}
\includegraphics[width=70mm]{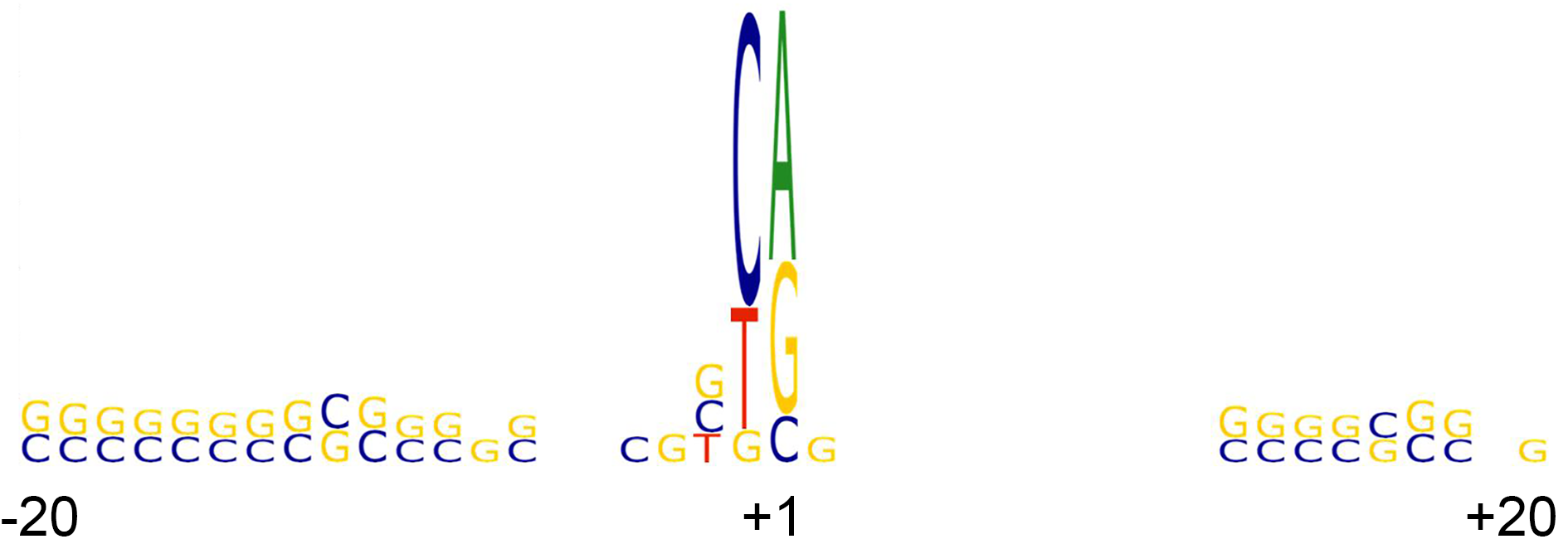}
\end{center}
\caption{Sequence logo of the region from -40 bp to +40 bp around the known TSS. The sequence logo demonstrates sequence conservation in the promoter initiator region and GC rich upstream  and downstream elements. }
\label{fig:logo_inr}
\end{figure}

To see the contributions of different nucleotides on different positions we computed so called feature saliency of all the sites in our testing set which is shown in Figure \ref{fig:sal}. At each position of the promoter region we replace a nucleotide with a new one for all the promoters and compute a new score. The rows represent different nucleotides which are used for replacement and columns show different positions inside the promoter region. If the new score on average increases it is represented by a red color. Score decrease is shown using a blue color. We can see that the largest effect on the score comes from the initiator region which has the most conserved motif in our data set (see Figure \ref{fig:logo_inr}). For the TSS position (+1), the most preferred nucleotides are \textit{A} and \textit{G}. If a promoter has an initiator then \textit{A} is the most frequent nucleotide at position +1, otherwise it is \textit{G}. This explains why \textit{A} and \textit{G} are red at position +1 in Figure \ref{fig:sal}. Changing nucleotides in region -30 bp to -23 bp  from the original ones to \textit{G} or \textit{C} reduces the score considerably. We know that a promoter region has more \textit{G} and \textit{C} nucleotides however the mentioned region has the TATA box that is why setting nucleotides to \textit{G} or \textit{C} has a negative effect on the score in this particular region. We also noticed two interesting trends. The most preferred nucleotide by our model before -100 bp is \textit{C}. However, after position +300 bp the score is clearly affected the most by the \textit{G} nucleotide.  

\subsection{Discussion}
As we have shown before, promoters with the TATA box can be predicted with a very small positional error; often the predicted TSS is exactly at the position of the true TSS. This is the result of strong TATA box fixating position of the promoter region. However, it is not the case for promoters without the TATA box for which the predicted positions have a normal distribution around the true TSS. For about 15\% of sequences in our testing set, predicted TSSs are farther than 100 bp from a true TSS. This problem can be partially explained  by occurrence of multiple TSSs in non-TATA promoters. Such promoters generate alternative gene isoforms that have  tissue or time specific expression. Many of such TSSs are not annotated in the promoter databases. It was shown in \cite{VoNgoc2017} that promoters have focused, dispersed, and mixed transcription. In dispersed transcription, there are many weak TSSs located at the region from -50 bp to +50 bp. This fact might be confusing for our deep learning model which results in a wider peak for a promoter (Figure \ref{fig:Landscape}).

\section{CONCLUSION}

While previously developed promoter prediction methods can  relatively accurately classify  promoter and non-promoter sequences, they fail to provide a good prediction when applied to study long genomic sequences. Due to potentially huge amount of tested locations they all have very low precision and generate a lot of false positives (often much more than the number of real promoters), that is limiting their usage in genome-scale studies.

In  this work we have proposed a novel training technique to overcome this issue. We used iterative training that focuses on instances that were misclassified by previous iterations
and builds our deep learning model that is able to eliminate the huge number of false positives. We analyzed different promoter regions to use as input for feature extraction
and chose optimal input location for our tool. Evaluation of our program performance and comparing it to the available promoter prediction tools demonstrated that PromID significantly outperforms other promoter finding programs. 

Many genes have non-coding exons and gene-finders can not provide the actual gene start and promoter position. Therefore programs for accurate computational identification of promoters are important  for revealing the gene structure and studying gene regulation. This work is a step towards this goal while we understand that this topic is open for further investigations on structure and functioning promoter regions.

\section{ACKNOWLEDGEMENTS}

\subsubsection{Conflict of interest statement.} None declared.
\newpage
\bibliographystyle{ieeetr}
\bibliography{PromID} 

\end{document}